# Transversal plasma resonance in a nonmagnetized plasma and possibilities of practical employment of it

F.F. Mende

E – mail: mende@mende.ilt.kharkov.ua

B.I. Verkin Institute for Low Temperature Physics and Engineering, NAS Ukraine,

47 Lenin Ave., Kharkov, 61164, Ukraiua

## Abstract

It is shown that in a nonmagnetized plasma, beside the longitudinal Langmuir resonance, there may also exist the transversal resonance. Both these resonance kinds are degenerated. Employment of the transversal resonance makes it possible to design resonators and filters, as well as powerful single-frequency lasers operating on the basis of collective oscillations of plasma.

## Introduction

Up to date it has been generally assumed that in a nonmagnetized plasma there may exist only one, so-called Langmuir resonance. The processes featuring it are associated with time-variation of the plasma density. As is well known, such a resonance is incapable of radiating electromagnetic waves in the longitudinal direction.

However, it is known that an explosion, for example of an atomic bomb, results in electromagnetic radiation in a very wide frequency band, up to radio-frequency band. While the radiation in a light band can be readily explained by radiation of separate atoms, the radiation in a radio-frequency band can be conditioned only by collective processes taking place in a hot plasma.

Revealing and describing of such processes is the objective of the present work.

## Electrodynamics of nonmagnetized plasma

Primarily we shall assume that we deal with a plasma medium in which there are no losses. For example, a superconductor is an ideal plasma medium where charge carriers may move without friction. In this case the equation of motion is

$$m\frac{d\vec{V}}{dt} = e\vec{E}, \tag{1}$$

where $m$ and $e$ are the electron mass and charge, respectively; $\vec{E}$ is the electric field strength, $\vec{V}$ is the velocity. Taking into account the current density

$$\vec{j} = ne\vec{V}, \tag{2}$$

we can obtain from Eq. (1)

$$\vec{j}_L = \frac{ne^2}{m}\int \vec{E}\, dt. \tag{3}$$

In Eqs. (2) and (3) $n$ is the specific charge density. Introducing the notion

$$L_k = \frac{m}{ne^2}, \tag{4}$$

we can write

$$\vec{j}_L = \frac{1}{L_k}\int \vec{E}\, dt. \tag{5}$$

Here $L_k$ is the kinetic inductivity of the medium. Its existence is based on the fact that a charge carrier has a mass and hence it possesses inertia properties.

For harmonic fields we have $\vec{E} = \vec{E}_0 \sin \omega t$ and Eq. (2.22) becomes

$$\vec{j}_L = -\frac{1}{\omega L_k} E_0 \cos \omega t. \tag{6}$$

Eqs. (5) and (6) show that $\vec{j}_L$ is the current through the inductance coil.

In this case the Maxwell equations take the following form

$$rot\,\vec{E} = -\mu_0 \frac{\partial \vec{H}}{\partial t},$$

$$rot\,\vec{H} = \vec{j}_C + \vec{j}_L = \varepsilon_0 \frac{\partial \vec{E}}{\partial t} + \frac{1}{L_k}\int \vec{E}\, dt, \tag{7}$$

where $\varepsilon_0$ and $\mu_0$ are the electric and magnetic inductivities in vacuum, $\vec{j}_C$ and $\vec{j}_L$ are the displacement and conduction currents, respectively. As was shown above, $\vec{j}_L$ is the inductive current.

Eq. (7) gives

$$rot\,rot\,\vec{H} + \mu_0\varepsilon_0 \frac{\partial^2 \vec{H}}{\partial t^2} + \frac{\mu_0}{L_k}\vec{H} = 0. \tag{8}$$

For time-independent fields, Eq. (8) transforms into the London equation

$$rot\,rot\,\vec{H} + \frac{\mu_0}{L_k}\vec{H} = 0, \tag{9}$$

where $\lambda_L^2 = \dfrac{L_k}{\mu_0}$ is the London depth of penetration.

As Eq. (7) shows, the inductivities of plasma (both electric and magnetic) are frequency – independent and equal to the corresponding parameters for vacuum. Besides, such plasma has another fundamental material characteristic – kinetic inductivity.

Eqs. (7) hold for both constant and variable fields. For harmonic fields $\vec{E} = \vec{E}_0 \sin \omega t$, Eq. (7) gives

$$rot\, \vec{H} = \left( \varepsilon_0 \omega - \dfrac{1}{L_k \omega} \right) \vec{E}_0 \cos \omega t. \qquad (10)$$

Taking the bracketed value as the specific susceptance $\sigma_x$ of plasma, we can write

$$rot\, \vec{H} = \sigma_X \vec{E}_0 \cos \omega t, \qquad (11)$$

where

$$\sigma_X = \varepsilon_0 \omega - \dfrac{1}{\omega L_k} = \varepsilon_0 \omega \left( 1 - \dfrac{\omega_\rho^2}{\omega^2} \right) = \omega\, \varepsilon^*(\omega), \qquad (12)$$

and $\varepsilon^*(\omega) = \varepsilon_0 \left( 1 - \dfrac{\omega_\rho^2}{\omega} \right)$, where $\omega_\rho^2 = \dfrac{1}{\varepsilon_0 L_k}$ is the plasma frequency.

Now Eq. (11) can be re-written as

$$rot\, \vec{H} = \omega\, \varepsilon_0 \left( 1 - \dfrac{\omega_\rho^2}{\omega^2} \right) \vec{E}_0 \cos \omega t, \qquad (13)$$

or

$$rot\, \vec{H} = \omega\, \varepsilon^*(\omega) \vec{E}_0 \cos \omega t. \qquad (14)$$

The $\varepsilon^*(\omega)$ –parameter is conventionally called the frequency-dependent electric inductivity of plasma. In reality however this magnitude includes simultaneously the electric inductivity of vacuum aid the kinetic inductivity of plasma. It can be found as

$$\varepsilon^*(\omega) = \dfrac{\sigma_X}{\omega}. \qquad (15)$$

It is evident that there is another way of writing $\sigma_X$

$$\sigma_X = \varepsilon_0 \omega - \dfrac{1}{\omega L_k} = \dfrac{1}{\omega L_k} \left( \dfrac{\omega^2}{\omega_\rho^2} - 1 \right) = \dfrac{1}{\omega L_k^*}, \qquad (16)$$

where

$$L_k^*(\omega) = \frac{L_k}{\left(\dfrac{\omega^2}{\omega_p^2} - 1\right)} = \frac{1}{\sigma_X \omega} \quad . \tag{17}$$

$L_k^*(\omega)$ written this way includes both $\varepsilon_0$ and $L_k$.

Eqs. (12) and (16) are equivalent, and it is safe to say that plasma is characterized by the frequency-dependent kinetic inductance $L_k^*(\omega)$ rather than by the frequency-dependent electric inductivity $\varepsilon^*(\omega)$.

Eq. (10) can be re-written using the parameters $\varepsilon^*(\omega)$ and $L_k^*(\omega)$

$$rot\ \vec{H} = \omega\ \varepsilon^*(\omega) \vec{E}_0 \cos\ \omega\ t \quad , \tag{18}$$

or

$$rot\ \vec{H} = \frac{1}{\omega\ L_k^*(\omega)} \vec{E}_0 \cos\ \omega\ t \quad . \tag{19}$$

Eqs. (18) and (19) are equivalent.

Thus, the parameter $\varepsilon^*(\omega)$ is not an electric inductivity though it has its dimensions. The same can be said about $L_k^*(\omega)$.

We can see readily that

$$\varepsilon^*(\omega) = \frac{\sigma_X}{\omega} \quad , \tag{20}$$

$$L_k^*(\omega) = \frac{1}{\sigma_X \omega} \quad . \tag{21}$$

These relations describe the physical meaning of $\varepsilon^*(\omega)$ and $L_k^*(\omega)$.

Of course, the parameters $\varepsilon^*(\omega)$ and $L_k^*(\omega)$ are hardly usable for calculating energy by the following equations

$$W_E = \frac{1}{2} \varepsilon\ E_0^2 \tag{22}$$

and

$$W_j = \frac{1}{2} L_k\ j_0^2 \ . \tag{23}$$

For this purpose the fotmula was devised in [1]:

$$W = \frac{1}{2} \cdot \frac{d[\omega\ \varepsilon^*(\omega)]}{d\omega} E_0^2 \quad . \tag{24}$$

Using Eq. (24), we can obtain

$$W_\Sigma = \frac{1}{2} \varepsilon_0 E_0^2 + \frac{1}{2} \cdot \frac{1}{\omega^2 L_k} E_0^2 = \frac{1}{2} \varepsilon_0 E_0^2 + \frac{1}{2} L_k\ j_0^2 \ . \tag{25}$$

The same result is obtainable from

$$W = \frac{1}{2} \cdot \frac{d\left[\frac{1}{\omega L_k^*(\omega)}\right]}{d\omega} E_0^2. \tag{26}$$

The parameters $\varepsilon^*(\omega)$ and $L_k^*(\omega)$ characterize completely the electrodynamics properties of plasma. The case

$$\begin{aligned}\varepsilon^*(\omega) &= 0 \\ L_k^*(\omega) &= \infty\end{aligned} \tag{27}$$

corresponds to the resonance of current.

It is shown below that under certain conditions this resonance can be transverse with respect to the direction of electromagnetic waves.
It is known that the Langmuir resonance is longitudinal. No other resonances have ever been detected in nonmagnetized plasma. Nevertheless, transverse resonance is also possible in such plasma, and its frequency coincides with that of the Langmuir resonance. To understand the origin of the transverse resonance, let us consider a long line consisting of two perfectly conducting planes (see Fig. 1).

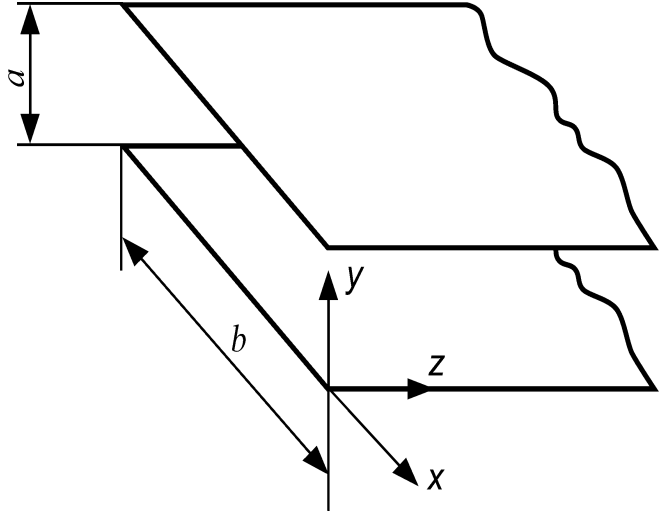

Fig.1. Two-conductor line consisting of two perfectly conducting planes.

First, we examine this line in vacuum.

If a d.c. voltage ($U$) source is connected to an open line the energy stored in its electric field is

$$W_{E\Sigma} = \frac{1}{2}\varepsilon_0 E^2 a\, b\, z = \frac{1}{2} C_{E\Sigma} U^2, \tag{28}$$

where $E = \dfrac{U}{a}$ is the electric field strength in the line, and

$$C_{E\Sigma} = \varepsilon_0 \frac{b\,z}{a} \tag{29}$$

is the total line capacitance. $C_E = \varepsilon_0 \dfrac{b}{a}$ is the linear capacitance and $\varepsilon_0$ is electric inductivities of the medium (plasma) in SI units (F/m).

The specific potential energy of the electric field is

$$W_E = \frac{1}{2}\varepsilon_0\, E^2 \;. \tag{30}$$

If the line is short-circuited at the distance $z$ from its start and connected to a d.c. current ($I$) source, the energy stored in the magnetic field of the line is

$$W_{H\Sigma} = \frac{1}{2}\mu_0\, H^2 a\, b\, z = \frac{1}{2} L_{H\Sigma}\, I^2 \;. \tag{31}$$

Since $H = \dfrac{I}{b}$, we can write

$$L_{H\Sigma} = \mu_0 \frac{a\,z}{b}, \tag{32}$$

where $L_{H\Sigma}$ is the total inductance of the line $L_H = \mu_0 \dfrac{a}{b}$ is linear inductance and $\mu_0$ is the inductivity of the medium (vacuum) in SI (H/m).

The specific energy of the magnetic field is

$$W_H = \frac{1}{2}\mu_0\, H^2 \;. \tag{33}$$

To make the results obtained more illustrative, henceforward, the method of equivalent circuits will be used along with mathematical description. It is seen that $C_{E\Sigma}$ and $L_{H\Sigma}$ increase with growing $z$. The line segment $dz$ can therefore be regarded as an equivalent circuit (Fig. 2a).

If plasma in which charge carriers can move free of friction is placed within the open line and then the current $I$, is passed through it, the charge carriers moving at a certain velocity start storing kinetic energy. Since the current density is

$$j = \frac{I}{b\,z} = n\,e\,V, \tag{34}$$

the total kinetic energy of all moving charges is

$$W_{k\Sigma} = \frac{1}{2} \cdot \frac{m}{n\,e^2} a\,b\,z\, j^2 = \frac{1}{2} \cdot \frac{m}{n\,e^2} \frac{a}{b\,z} I^2 \;. \tag{35}$$

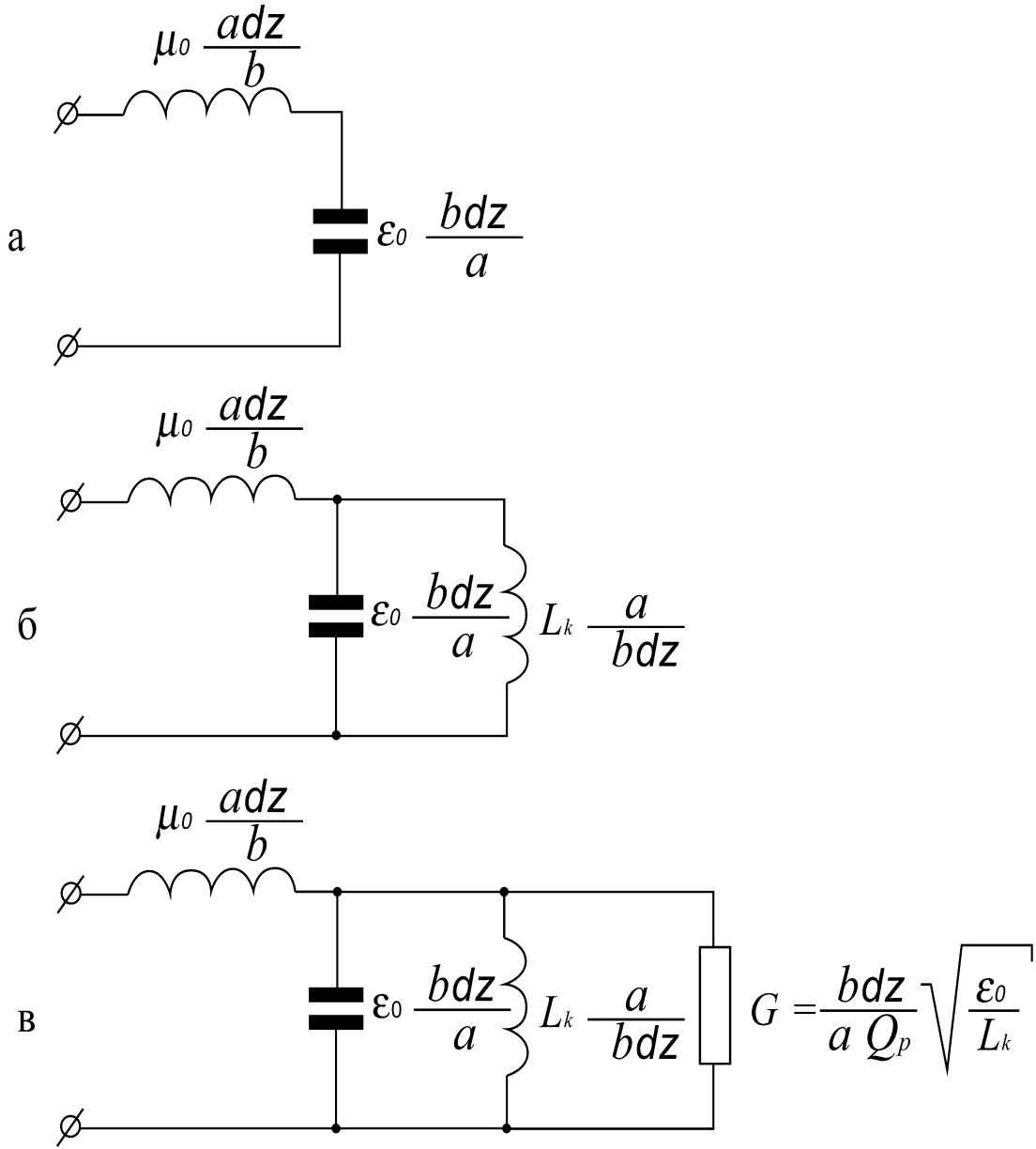

Fig. 2. а. Equivalent circuit of the two-conductor line segment;
  б. Equivalent circuit of the two-conductor line segment containing nondissipative plasma;
  в. Equivalent circuit of the two-conductor line segment containing dissipative plasma.

On the other hand,

$$W_{k\Sigma} = \frac{1}{2} L_{k\Sigma} I^2 , \qquad (36)$$

where $L_{k\Sigma}$ is the total kinetic inductance of the line. Hence,

$$L_{k\Sigma} = \frac{m}{n\,e^2} \cdot \frac{a}{b\,z} \quad . \tag{37}$$

Thus, the magnitude

$$L_k = \frac{m}{n\,e^2} \tag{38}$$

corresponding kinetic inductivity of the medium.

Earlier, we introduced this magnitude by another way (see Eq. (4)). Eq. (38) corresponds to case of uniformly distributed d.c. current.

As we can see from Eq. (37), $L_{H\Sigma}$, unlike $C_{E\Sigma}$ and $L_{k\Sigma}$, decreases when $z$ grows. This is clear physically because the number of parallel-connected inductive elements increases with growing $z$. The equivalent circuit of the line with non-dissipative plasma is shown in Fig. 26. The line itself is equivalent to a parallel lumped circuit:

$$C = \frac{\varepsilon_0 b\,z}{a} \quad \text{and} \quad L = \frac{L_k\,a}{b\,z}. \tag{39}$$

It is however obvious from calculation that the resonance frequency is absolutely independent of whatever dimension. Indeed,

$$\omega_\rho^2 = \frac{1}{C\,L} = \frac{1}{\varepsilon_0 L_k} = \frac{n\,e^2}{\varepsilon_0 m} \quad . \tag{40}$$

This brings us to a very interesting result: the resonance frequency of the macroscopic resonator is independent of its size. It may seem that we are dealing here with the Langmuir resonance because the obtained frequency corresponds exactly to that of the Langmuir resonance. We however know that the Langmuir resonance characterizes longitudinal waves. The wave propagating in the phase velocity in the $z$-direction is equal to infinity and the wave vector is $\vec{k}_z = 0$, which corresponds to the solution of Eqs. (7) for a line of pre-assigned configuration (Fig.1). Eqs. (8) give a well-known result. The wave number is

$$k_z^2 = \frac{\omega^2}{c^2}\left(1 - \frac{\omega_\rho^2}{\omega^2}\right). \tag{41}$$

The group and phase velocities are

$$V_g^2 = c^2\left(1 - \frac{\omega_\rho^2}{\omega^2}\right), \tag{42}$$

$$V_F^2 = \frac{c^2}{\left(1 - \frac{\omega_\rho^2}{\omega^2}\right)}, \tag{43}$$

where $c = \left(\frac{1}{\mu_0 \varepsilon_0}\right)^{1/2}$ is the velocity of light in vacuum.

For the plasma under consideration, the phase velocity of the electromagnetic wave is equal to infinity. Hence, the distribution of the fields and currents over the line is uniform at each instant of time and independent of the $z$-coordinate. This implies that, on the one hand, the inductance $L_{H\Sigma}$ has no effect on the electrodynamic processes in the line and, on the other hand, any two planes can be used instead of conducting planes to confine plasma above and below.

Eqs. (41), (42) and (43) indicate that we have transverse resonance with an infinite $Q$-factor. The fact of transverse resonance, i.e. different from the Langmuir resonance, is most obvious when the $Q$-factor is not equal to infinity. Then $k_z \neq 0$ and the transverse wave is propagating in the line along the direction perpendicular to the movement of charge carriers. True, we started our analysis with plasma confined within two planes of a long line, but we have thus found that the presence of such resonance is entirely independent of the line size, i.e. this resonance can exist in an infinite medium. Moreover, in infinite plasma transverse resonance can coexist with the Langmuir resonance characterizing longitudinal waves. Since the frequencies of these resonances coincide, both of them are degenerate. Earlier, the possibility of transverse resonance was not considered. To approach the problem more comprehensively, let us analyze the energy processes in loss-free plasma.

The characteristic resistance of plasma determining the relation between the transverse components of electric and magnetic fields can be found from

$$Z = \frac{E_y}{H_x} = \frac{\mu_0 \omega}{k_z} = Z_0 \left(1 - \frac{\omega_\rho^2}{\omega^2}\right)^{-1/2}, \tag{44}$$

where $Z_0 = \sqrt{\frac{\mu_0}{\varepsilon_0}}$ is the characteristic resistance in vacuum.

The obtained value of $Z$ is typical for transverse electromagnetic waves in waveguides. When $\omega \to \omega_\rho$, $Z \to \infty$, and $H_x \to 0$. At $\omega > \omega_\rho$, both the electric and magnetic field components are present in plasma. The specific energy of the fields is

$$W_{E,H} = \frac{1}{2} \varepsilon_0 E_{0y}^2 + \frac{1}{2} \mu_0 H_{0x}^2. \tag{45}$$

Thus, the energy accumulated in the magnetic field is $\left(1 - \frac{\omega_\rho^2}{\omega^2}\right)$ times lower than that in the electric field. This traditional electrodynamic analysis is however not complete because it disregards one more energy component – the kinetic energy of charge carriers. It turns out that in addition to the electric and magnetic waves carrying electric and magnetic energy, there is one more wave in plasma – the kinetic wave carrying the kinetic energy of charge carriers. The specific energy of this wave is

$$W_k = \frac{1}{2} L_k j_0^2 = \frac{1}{2} \cdot \frac{1}{\omega^2 L_k} E_0^2 = \frac{1}{2} \varepsilon_0 \frac{\omega_p^2}{\omega^2} E_0^2 \ . \tag{46}$$

The total specific energy thus amounts to

$$W_{E,H,j} = \frac{1}{2} \varepsilon_0 E_{0y}^2 + \frac{1}{2} \mu_0 H_{0x}^2 + \frac{1}{2} L_k j_0^2 \ . \tag{47}$$

Hence, to find the total specific energy accumulated in unit volume of plasma, it is not sufficient to allow only for the fields $E$ and $H$.

At the point $\omega = \omega_p$

$$\begin{aligned} W_H &= 0 \\ W_E &= W_k \ , \end{aligned} \tag{48}$$

i.e. there is no magnetic field in the plasma, and the plasma is a macroscopic electromechanical cavity resonator of frequency $\omega_p$..

At $\omega > \omega_p$ the wave propagating in plasma carries three types of energy – magnetic, electric and kinetic. Such wave can therefore be-called magnetoelectrokinetic. The kinetic wave is a current-density wave $\vec{j} = \frac{1}{L_k} \int \vec{E} \, dt$. It is shifted by $\pi/2$ with respect to the electric wave.

Up to now we have considered a physically unfeasible case with no losses in plasma, which corresponds to infinite $Q$-factor of the plasma resonator. If losses occur, no matter what physical processes caused them, the $Q$-factor of the plasma resonator is a final quantity. For this case the Maxwell equations become

$$\begin{aligned} \text{rot } \vec{E} &= -\mu_0 \frac{\partial \vec{H}}{\partial t}, \\ \text{rot } \vec{H} &= \sigma_{p.ef} \vec{E} + \varepsilon_0 \frac{\partial \vec{E}}{\partial t} + \frac{1}{L_k} \int \vec{E} \, dt. \end{aligned} \tag{49}$$

The term $\sigma_{p.ef} \vec{E}$ allows for the loss, and the index *ef,* near the active conductivity emphasizes that we are interested in the fact of loss and do not care of its mechanism. Nevertheless, even though we do not try to analyze the physical mechanism of loss, we should be able at least to measure $\sigma_{p.ef}$.

For this purpose, we choose a line segment of the length $z_0$, which is much shorter than the wavelength in dissipative plasma. This segment is equivalent to a circuit with the following lumped parameters

$$C = \varepsilon_0 \frac{b \, z_0}{a}, \tag{50}$$

$$L = L_k \frac{d}{b\, z_0}, \tag{51}$$

$$G = \sigma_{p.ef} \frac{b\, z_0}{a}, \tag{52}$$

where $G$ is the conductance parallel to $C$ and $L$.

The conductance $G$ and the $Q$-factor of this circuit are related as

$$G = \frac{1}{Q_\rho} \sqrt{\frac{C}{L}}. \tag{53}$$

Taking into account Eqs. (50) – (52), we obtain from Eq. (53)

$$\sigma_{p.ef} = \frac{1}{Q_\rho} \sqrt{\frac{\varepsilon_0}{L_k}}. \tag{54}$$

Thus, $\sigma_{p.ef.}$ can be found by measuring the basic $Q$-factor of the plasma resonator.

Using Eqs. (54) and (49), we obtain

$$rot\, \vec{E} = -\mu_0 \frac{\partial \vec{H}}{\partial t},$$

$$rot\, \vec{H} = \frac{1}{Q_\rho} \sqrt{\frac{\varepsilon_0}{L_k}} \vec{E} + \varepsilon_0 \frac{\partial \vec{E}}{\partial t} + \frac{1}{L_k} \int \vec{E}\, dt. \tag{55}$$

The equivalent circuit of this line containing dissipative plasma is shown in Fig. 2b.

Lot us consider the solution of Eqs. (55) at the point $\omega = \omega_p$. Since

$$\varepsilon_0 \frac{\partial \vec{E}}{\partial t} + \frac{1}{L_k} \int \vec{E}\, dt = 0, \tag{56}$$

we obtain

$$rot\, \vec{E} = -\mu_0 \frac{\partial \vec{H}}{\partial t},$$

$$rot\, \vec{H} = \frac{1}{Q_P} \sqrt{\frac{\varepsilon_0}{L_k}} \vec{E}. \tag{57}$$

The solution of these equations is well known. If there is interface between vacuum and the medium described by Eqs. (57), the surface impedance of the medium is

$$Z = \frac{E_{tg}}{H_{tg}} = \sqrt{\frac{\omega_p \mu_0}{2\sigma_{p.ef.}}}(1+i), \tag{58}$$

where $\sigma_{p.ef} = \frac{1}{Q_p}\sqrt{\frac{\varepsilon_0}{L_k}}$.

There is of course some uncertainty in this approach because the surface impedance is dependent on the type of the field-current relation (local or non-local). Although the approach is simplified, the qualitative results are quite adequate. True, a more rigorous solution is possible.

The wave propagating deep inside the medium decreases by the law $e^{-\frac{z}{\delta_{ef}}} \cdot e^{-i\frac{z}{\delta_{ef}}}$. In this case the phase velocity is

$$V_F = \omega\, \sigma_{p.ef}, \tag{59}$$

where $\delta_{p.ef}^2 = \frac{2}{\mu_0 \omega_p \sigma_{p.ef}}$ is the effective depth of field penetration in the plasma.

The above relations characterize the wave process in plasma. For good conductors we usually have $\frac{\sigma_{ef}}{\omega\,\varepsilon_0} \gg 1$. In such a medium the wavelength is

$$\lambda_g = 2\pi\delta. \tag{60}$$

I.e. much shorter than the free-space wavelength. Further on we concentrate on the case $\lambda_g \gg \lambda_0$ at the point $\omega = \omega_p$, i.e. $V_F\big|_{\omega=\omega p} \gg c$.

## Practical aspects

Plasma can be used first of all to construct a macroscopic single-frequency cavity for development of a new class of electrokinetic plasma lasers. Such cavity can also operate as a band-pass filter.

At high enough $Q_p$ the magnetic field energy near the transverse resonance is considerably lower than the kinetic energy of the current carriers and the electrostatic field energy. Besides, under certain conditions the phase velocity can much exceed the velocity of light. Therefore, if we want to excite the transverse plasma resonance, we can put

$$rot\,\vec{E} \cong 0,$$

$$\frac{1}{Q_p}\sqrt{\frac{\varepsilon_0}{L_k}}\vec{E} + \varepsilon_0 \frac{\partial\vec{E}}{\partial t} + \frac{1}{L_k}\int \vec{E}\,dt = \vec{j}_{CT}, \tag{61}$$

where $\vec{j}_{CT}$ is the extrinsic current density.

Differentiating Eq. (61) over time and dividing it by $\varepsilon_0$ obtain

$$\omega_p^2 \vec{E} + \frac{\omega_p}{Q_p} \cdot \frac{\partial \vec{E}}{\partial t} + \frac{\partial^2 \vec{E}}{\partial t^2} = \frac{1}{\varepsilon_0} \cdot \frac{\partial \vec{j}_{CT}}{\partial t}. \tag{62}$$

Integrating Eq. (62) over the surface normal to the vector $\vec{E}$ and taking $\Phi_E = \int \vec{E}\, d\vec{S}$, we have

$$\omega_p^2 \Phi_E + \frac{\omega_p}{Q_p} \cdot \frac{\partial \Phi_E}{\partial t} + \frac{\partial^2 \Phi_E}{\partial t^2} = \frac{1}{\varepsilon_0} \cdot \frac{\partial I_{CT}}{\partial t}, \tag{63}$$

where $I_{CT}$ is the extrinsic current.

Eq. (63) is the harmonic oscillator equation whose right-hand side is typical of two-level lasers [2]. If there is no excitation source, we have a "cold". Laser cavity in which the oscillation damping follows the exponential law

$$\Phi_E(t) = \Phi_E(0) e^{i\omega_p t} \cdot e^{-\frac{\omega_p}{2Q_p} t}, \tag{64}$$

i.e. the macroscopic electric flow $\Phi_E(t)$ oscillates at the frequency $\omega_p$. The relaxation time can be round as

$$\tau = \frac{2Q_P}{\omega_P}. \tag{65}$$

If this cavity is excited by extrinsic currents, the cavity will operate as a band-pass filter with the pass band $\Delta\omega = \frac{\omega_p}{2Q_p}$.

Transverse plasma resonance offers another important application – it can be used to heat plasma. High-level electric fields and, hence, high change-carrier energies can be obtained in the plasma resonator if its $Q$-factor is high, which is achievable at low concentrations of plasma. Such cavity has the advantage that the charges attain the highest velocities far from cold planes. Using such charges for nuclear fusion, we can keep the process far from the cold elements of the resonator.

Let us pay special attention on that, in this case, when the plasma is bounded (similar to a case of two-wire circuit), it is possible to perform selective heating of the plasma. We keep in mind that such a resonator will resonate separately both at frequencies of electron and ion plasma resonance. This is associated with that, in a case of bounded plasma, at frequencies that are considerably lower than ones of electron plasma resonance, the electrons will practically instantly follow the field. This will result in that the electrons synchronously following the field will cause a certain gradient of the charge density and will form in such a way some compensating field directed against the primary field. The strength of this compensating field will be independent on the frequency in the frequency band mentioned.

In spite of that the electrons will exert a stabilizing action, and the field strength between the pates will be less than in the case of electron absence, yet an alternating field between the plates will exists. In the given case, a low-frequency resonance will occur in that case when the external frequency will coincide with the ion resonance frequency. Note that the possibility of separate heating of the electron and ion plasma components has been absent till now. There was only the possibility of heating of the electron component, whereas for implementation on thermonuclear synthesis just heating of the ion component is needed. When considering this issue we imply that the conducting plates themselves of the circuit are conductively isolated from the volume occupied by the plasma.

Such plasma resonator can be matched easily to the communication line. Indeed, the equivalent resistance of the resonator at the point $\omega = \omega_p$ is

$$R_{экв} = \frac{1}{G} = \frac{a\, Q_P}{b\, z} \sqrt{\frac{L_k}{\varepsilon_0}}. \tag{66}$$

The communication lines of sizes $a_L$ and $b_L$ should be connected to the cavity either through a smooth junction or in a stepwise manner. If $b = b_L$, the matching requirement is

$$\frac{a_L}{b_L} \sqrt{\frac{\mu_0}{\varepsilon_0}} = \frac{a\, Q_p}{b\, z_0} \sqrt{\frac{L_k}{\varepsilon_0}}, \tag{67}$$

$$\frac{a\, Q_p}{a_L z_0} \sqrt{\frac{L_k}{\mu_0}} = 1. \tag{68}$$

It should be remembered that the choice of the resonator length $z_0$ must comply with the requirement $z_0 \ll \lambda_g \big|_{\omega=\omega p}$.

Development of devices based on plasma resonator can require coordination of the resonator and free space. In this case the following condition is important:

$$\sqrt{\frac{\mu_0}{\varepsilon_0}} = \frac{a\, Q_p}{b\, z_0} \sqrt{\frac{L_k}{\varepsilon_0}}, \tag{69}$$

or

$$\frac{a\, Q_p}{b\, z_0} \sqrt{\frac{L_k}{\mu_0}} = 1. \tag{70}$$

.

## Discussion of results

We have found that $\varepsilon(\omega)$ is not dielectric inductivity permittivity. Instead, it includes two frequency-independent parameters $\varepsilon_0$ and $L_k$. What is the reason for the physical misunderstanding of the parameter $\varepsilon(\omega)$? This occurs first of all because for the case of plasma the $\frac{1}{L_k} \int \vec{E}\, dt$ - type term is not explicitly present in the second Maxwell equation.

There is however another reason for this serious mistake in the present-day physics [1] as an example. This study states that there is no difference between dielectrics and conductors at very high frequencies. On this basis the authors suggest the existence of a polarization vector in conducting media and this vector is introduced from the relation

$$\vec{P} = \Sigma\, e\, \vec{r}_m = n\, e\, \vec{r}_m, \tag{71}$$

where $n$ is the charge carrier density, $\vec{r}_m$ is the current charge displacement. This approach is physically erroneous because only bound charges can polarize and form electric dipoles when the external field overcoming the attraction force of the bound charges accumulates extra electrostatic energy in the dipoles. In conductors the charges are not bound and their displacement would not produce any extra electrostatic energy. This is especially obvious if we employ the induction technique to induce current (i.e. to displace charges) in a ring conductor. In this case there is no restoring force to act upon the charges, hence, no electric polarization is possible. In [1] the polarization vector found from Eq. (71) is introduced into the electric induction of conducting media

$$\vec{D} = \varepsilon_0\, \vec{E} + \vec{P}, \tag{72}$$

where the vector $\vec{P}$ of a metal is obtained from Eq. (71), which is wrong.

Since

$$\vec{r}_m = -\frac{e^2}{m\, \omega^2}\vec{E}, \tag{73}$$

for free carriers, then

$$\vec{P}^*(\omega) = -\frac{n\, e^2}{m\, \omega^2}\vec{E}, \tag{74}$$

for plasma, and

$$\vec{D}^*(\omega) = \varepsilon_0\, \vec{E} + \vec{P}^*(\omega) = \varepsilon_0\left(1 - \frac{\omega_p^2}{\omega^2}\right)\vec{E}. \tag{75}$$

Thus, the total accumulated energy is

$$W_\Sigma = \frac{1}{2}\varepsilon_0\, E^2 + \frac{1}{2}\cdot\frac{1}{L_k\, \omega^2}E^2. \tag{76}$$

However, the second term in the right-hand side of Eq. (76) is the kinetic energy (in contrast to dielectrics for which this term is the potential energy). Hence, the electric induction vector $D^*(\omega)$ does not correspond to the physical definition of the electric induction vector.

The physical meaning of the introduced vector $\vec{P}^*(\omega)$ is clear from

$$\vec{P}^*(\omega) = \frac{\sigma_L}{\omega}\vec{E} = \frac{1}{L_k\, \omega^2}\vec{E}. \tag{77}$$

The interpretation of $\varepsilon(\omega)$ as frequency-dependent inductivity has been harmful for correct understanding of the real physical picture (especially in the educational processes). Besides, it has drawn away the researchers attention from some physical phenomena in plasma, which first of all include the transverse plasma resonance and three energy components of the magnetoelectrokinetic wave propagating in plasma.

## Conclusions

The author is indebted to V.D.Fil for helpful discussions, to N.P.Mende for their assistance in preparation of this manuscript.